# A silicon photonics waveguide-coupled colloidal quantum dot photodiode sensitive beyond 1.6 µm

CHAO PANG[1,2,3,*], YU-HAO DENG[2,3], EZAT KHERADMAND[2,3], LUIS MORENO HAGELSIEB[4], YUJIE GUO[1,2], DAVID CHEYNS[4], PIETER GEIREGAT[2,3], ZEGER HENS[2,3], DRIES VAN THOURHOUT[1,2]

[1] *Photonics Research Group, Ghent University - imec, 9052 Gent, Belgium*

[2] *NB Photonics, Ghent University, 9052 Gent, Belgium*

[3] *Physics and Chemistry of Nanostructures Group, Ghent University, 9000 Gent, Belgium*

[4] *IMEC, 3001 Leuven, Belgium*

*Chao.Pang@Ugent.be*

**Abstract:** Silicon photonics faces a persistent challenge in extending photodetection capabilities beyond the 1.6 µm wavelength range, primarily due to the lack of appropriate epitaxial materials. Colloidal quantum dots (QDs) present a promising solution here, offering distinct advantages such as infrared wavelength tunability, cost-effectiveness, and facile deposition. Their unique properties position them as a potential candidate for enabling photodetection in silicon photonics beyond the conventional telecom wavelength, thereby expanding the potential applications and capabilities within this domain. In this study, we have successfully integrated lead sulfide (PbS) colloidal quantum dot photodiodes (QDPDs) onto silicon waveguides using standard process techniques. The integrated photodiodes exhibit a remarkable responsivity of 1.3 A/W (with an external quantum efficiency of 74.8%) at a wavelength of 2.1 µm, a low dark current of only 106 nA and a bandwidth of 1.1 MHz under a -3 V bias. To demonstrate the scalability of our integration approach, we have developed a compact 8-channel spectrometer incorporating an array of QDPDs. This achievement marks a significant step toward realizing a cost-effective photodetector solution for silicon photonics, particularly tailored for a wide range of sensing applications around the 2 µm wavelength range.

## 1. INTRODUCTION

Silicon photonics is renowned for its capability to miniaturize complex bulk systems into cost-effective and robust chips, which find applications in fields ranging from data/tele communications to sensing[1–3]. Despite the widespread use of silicon photonics, current applications predominantly use light with wavelengths shorter than 1.6 µm. This limitation is not a choice.

Numerous applications, including environmental monitoring, medical diagnostics, and industrial sensing, would profit from photonic chips operating at longer wavelengths [4,5]. Such applications are, however, hampered by the lack of a scalable and cost-effective integrated technology for photodetection beyond 1.6 µm.

Different strategies are being explored to extend the photodetection wavelength range of silicon photonics. A commonly used technique involves wafer or die bonding of III-V materials onto silicon, utilizing GaSb based compounds[6–9]. However, this method requires large-area planarized surfaces and relies on costly III-V epitaxy and bonding technology. Also two-dimensional (2D) materials like graphene and black phosphorus are extensively investigated for on-chip photodetection beyond 1.6 µm[10]. The promise of these materials lies in the favorable substrate compatibility and broad absorption range, yet scalability poses a substantial hurdle for the practical integration of 2D photodetectors. Moreover, the performance of these photodetectors needs further improvement, for example to enhance dark current performance and responsivity[11,12]. Alternatively, the direct deposition on Si of GeSn alloys[13–16], III-V quantum dots[17], or poly-crystalline tellurium thin films[18] can result in cost-effective photodetectors with straightforward fabrication processes. However, the current implementations of these methods result in photodetectors with high dark current and low responsivity, necessitating further development.

To address the challenge of cost-effective photodetection beyond the telecom range in silicon photonics, colloidal quantum dots (QDs) offer a promising solution. First of all, QDs have economical advantage in both material synthesis and deposition processes. Synthesized through wet-chemical methods, they exhibit significantly lower material cost (10\$ to 60\$ per gram[19]), and can be deposited as thin films on 200 mm silicon wafers through, for example, spin-coating or spray coating. The excellent compatibility of QDs with various substrates make QDs particularly well-suited for heterogeneous material integration, in stark contrast with more traditional III-V epitaxial semiconductors. Moreover, the absorption spectrum of QDs can be readily tuned across the infrared by adjusting the QD size and utilizing different materials. Photodetectors made from PbS[20–23], HgTe[24,25], $Ag_2Se$[26,27] or InAs[28] QDs have all shown sensitivity for short-wave and mid-wave infrared light.

Given the appealing features of QDs, interest is growing in integrating QD-based emitters and detectors on photonic chips. In particular for visible light, electrically pumped light emitting devices and optically pumped lasers have been demonstrated by combining CdSe-based QDs with silicon nitride (SiN) photonic chips[29–33]. More recently, waveguide-coupled QD-based photodiodes operating at around 1.3 µm were demonstrated on the same SiN platform. This achievement underscored the potential of QDs as scalable photodetectors within photonic integrated circuits (PICs)[34]. For wavelengths beyond 1.6 µm, recent developments include the demonstration of plasmonic HgTe-based photoconductors on silicon waveguides[35] operating at 2.3 µm with a responsivity of 23 mA/W. On the other hand, PbS QD-based photodiodes (QDPDs) sensitive up to 2.1 µm with a responsivity of 0.385 A/W were achieved by sandwiching the QD film between a NiO p-type and a ZnO n-type contact[36]. Such reports point towards to promise of using integrated QDPDs for photodetection beyond 1.6 µm, provided that low dark currents and high responsivity can be realized in integration-ready QDPD stacks.

In this work, we report on a silicon photonics, waveguide-coupled QD photodiode (WG-QDPD) that features a spectral response extending beyond 1.6 µm. Leveraging an optimized QDPD design and a proven process flow[34], we demonstrate WG-QDPDs exhibiting a responsivity of 1.3 A/W at 2.1 µm, a low dark current of 106 nA, and a bandwidth of 1.1 MHz. Moreover, we show that our WG-QDPD exhibits a low noise equivalent power (NEP) of 0.15 pW/$\sqrt{Hz}$, due to its high responsivity and low dark current, which is attractive for low-noise, weak signal detection. We demonstrate the reliability of the integration approach by realizing a compact on-chip spectrometer that features an array of 8 WG-QDPDs and operates in the spectral window between 2.063 and 2.135 µm.

## 2. RESULTS AND DISCUSSION

*2.1 QDPDs with wavelength response up to 2.1 µm*

Before integrating QDPDs on waveguides, we optimized the material stack using dummy surface-illuminated QDPDs fabricated on glass-ITO substrates. Our approach involved adapting the p-i-n structure developed to form 1.3 µm WG-QDPDs for PbS QDs with a band gap transition at 2.1 µm [34]. All PbS QDs were synthesized by reacting lead oleate with a substituted thiourea, where different QD sizes were obtained by using differently substituted thioureas in agreement with literature[37]. The resulting PbS QDs have a surface terminated by lead oleate. More details on the synthesis and the materials characteristics can be found in **APPENDIX A**.

As shown in Figure 1a, we made QDPD stacks consisting of a 100 nm thick ITO transparent bottom electrode, a 30 nm thick sol-gel ZnO film functioning as an electron transport and hole blocking layer, a multilayer QD stack and an Au top electrode. Within the QD film, the photosensitive layer consisted of QDs with a band-gap transition at 2.1 µm (**APPENDIX A Figure 7**). In a first design (Structure A in Figure 1a), this film – formed by spin coating – was exposed to a methanol solution containing 10 mg/mL tetra-n-butylammonium iodide (TBAI) for 30 s. This process removes as-synthesized surface ligands, thereby turning the QD film *n*-type and enhancing the electron mobility. As the hole transport and electron blocking layer, we employed a film of PbS QDs featuring a 0.94 µm band-gap transition. This film was similarly exposed to a 0.01 vol% ethanedithiol (EDT) in methanol solution, which leads to *p*-type doping[38,39], and a gold contact was evaporated on top.

However, as shown in Figure 1b (red line), this stack showed little rectification, and a dark current density of nearly 10 mA/cm$^2$ at -2 V reverse bias, which is more than two orders higher than QDPDs using PbS QDs with 1.3 µm bandgap transition as photosensitive layers (**APPENDIX B Figure 8**). In an attempt to reduce the dark current, we explored two additional QDPD stacks; a first in which the film of 2.1 µm PbS QDs was sandwiched between layers of 1.3 µm PbS QDs, which we also treated using TBAI and a second where the 1.3 µm PbS QDs were only included between the 2.1 and 0.94 µm PbS QD films, see Structure B and C in Figure 1a. Interestingly, structure B showed a strongly enhanced rectification and a 10-fold reduction of the dark current as compared to the initial stack, while structure C only featured a minor reduction of the dark current. This difference indicates that the primary source of dark current is the leakage of charge carriers between the 2.1 µm PbS QDs and the ZnO contact, possibly assisted through trap states at the ZnO surface[20], for which the 1.3 µm PbS QD film provides an additional barrier.

The optimized QDPD with Structure B was characterized with a 2.1 µm laser that had a power of 870 µW and a peak power density of 220 mW/cm$^2$. Under a -3 V bias, the QDPD shows a dark current of 50.3 µA (dark current density of 2.8 mA/cm² for a device area of 1.77 mm$^2$) and a responsivity of 0.19 A/W (external quantum efficiency, EQE, of 11.2%), as illustrated in Figure 1c. This relatively low EQE probably reflects the limited absorption of the vertically incident light in the thin layer of 2.1 µm PbS QDs, and the relatively low transmittance of ITO at 2.1 µm[36]. The 1.3 µm PbS-TBAI layer between the 2.1 µm PbS-TBAI and PbS-EDT layers should be n-doped[38], posing challenges for hole transport and potentially leading to an increase in the series resistance of the QDPD. However, the I-V characteristics at forward bias did not show degradation compared to Structure A (Figure 1b). Possibly, this layer is turned p-type during the subsequent EDT ligand exchange process for the 0.94 µm PbS QD layer on top. EDT, known for its high reactivity compared to TBAI[40], can soak the 1.3 µm PbS-TBAI layer underneath, resulting in a change of doping to p-type. To validate this hypothesis, QDPDs with Structure B were fabricated, but the ligand exchange of the top 1.3 µm PbS QDs was altered to EDT instead of TBAI. QDPDs with either TBAI or EDT ligand exchange on the 1.3 µm QDs exhibited nearly identical dark and light I-V characteristics (**APPENDIX B Figure 9**), thus confirming our above assumption. The spectral response of the QDPD with structure B was measured by sweeping the laser wavelength, showing photo response extending up to 2.2 µm (**APPENDIX B Figure 10**), consistent with the absorption spectrum of the QDs. More details regarding the fabrication of the QDPD can be found in **APPENDIX C.**

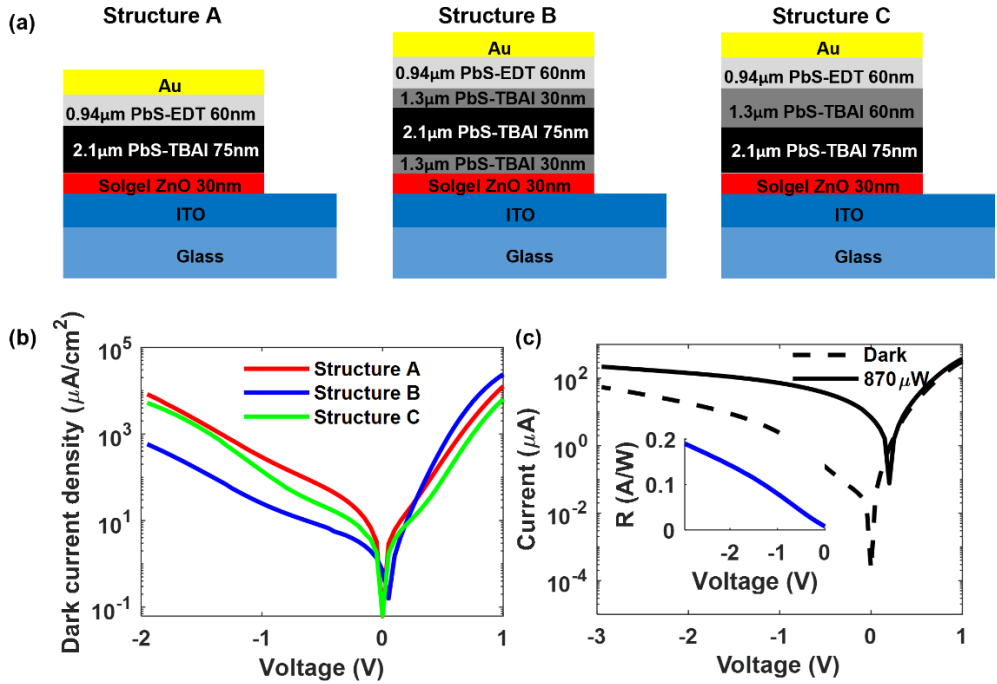

Fig. 1. Structure and characterization of QDPD. (a) Vertical QDPD structures with surface illumination. (b) Influence of the interface layer on dark current density. (c) Dark current and photocurrent of Structure B QDPD, with an area of 1.77 mm$^2$. The photocurrent was measured with 2.1 µm laser gaussian beam illuminated from the bottom glass side, with a peak power density of 220 mW/cm$^2$. Inset: responsivity of QDPD vs. bias voltage.

*2.2 Integration of QDPDs on waveguides*

Figure 2a-e depicts the process flow used to integrate the QDPD on waveguides. In a first step, waveguides were defined on silicon-on-insulator (SOI) substrates with a 2 µm buried oxide and a 220 nm silicon layer, employing a 70 nm shallow etch. To alleviate the power saturation problem of WG-QDPDs, as observed in previous demonstration at shorter wavelength on the SiN platform[34], we defined 30 µm wide waveguides. This adjustment serves to lower the optical power density in the QD film. Next, the waveguide was planarized by flowable oxide (hydrogen silsesquioxane, HSQ), followed by annealing at 400 °C in N$_2$ to increase the physical and chemical stability of HSQ. This step resulted in a 45 nm top cladding on the waveguides, shown in light blue in Figure 2a. As outlined in Figure 2b-e, the QDPD stack was then integrated on top, starting with sputtering of an 18 nm thick ITO layer as the bottom electrode. The ITO had a sheet resistance of 173 Ω/□, and the layer thickness was reduced as compared to the dummy surface-illuminated QDPD structure in order to minimize optical losses in the ITO and maximize light absorption in the PbS QD film. To further reduce optical loss, the ITO layer was restricted to the designated

area for photodetection using HCl-based wet etching. Next, we deposited the ZnO layer using sol-gel chemistry and patterned it using a dilute HCl solution. Complete coverage of ITO with ZnO is crucial to prevent direct contact between ITO and the subsequently deposited QD film, as this could lead to a significant leakage current. Ti/Au (20 nm / 100 nm) was placed on the side of the waveguides as the n-contact pad using a lift-off technique. The same QD stacks with structure B, as we used in the surface-illuminated QDPD, were deposited layer by layer and lifted off simultaneously to achieve the desired pattern. PMMA resist was used for its excellent chemical compatibility with the methanol solvent involved during QDs deposition. Finally, 80 nm Au was deposited on top as p-contact pads through thermal evaporation and lift-off. The cross-section of the fabricated WG-QDPD is shown in Fig. 2f. More details on the integration process can be found in **APPENDIX D.**

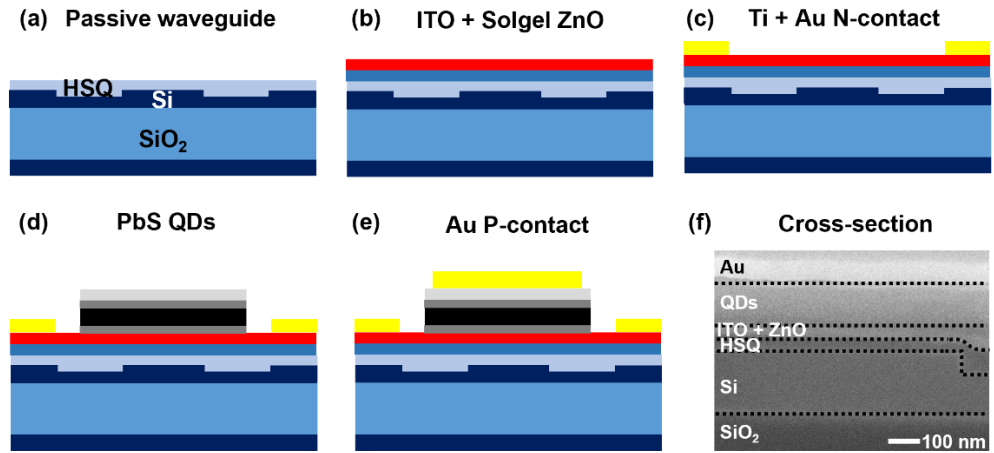

Fig. 2. Integration steps to fabricate WG-QDPD. (a) Si rib waveguide patterning and planarization. (b) ITO sputtering and patterning with wet etching, followed by ZnO deposition with sol-gel method and patterning with wet etching. (c) N-contact metal patterning with liftoff. (d) QD film patterning with PMMA resist and liftoff. (e) P-contact metal patterning with PMMA resist and liftoff. (f) Cross-section of fabricated WG-QDPD.

## 2.3 Characterization of WG-QDPD

We characterized the WG-QDPDs using a 2.1 µm laser, employing grating couplers for convenient light coupling into and out of the waveguide. The optical power in the waveguide was calibrated by measuring the coupling efficiency of the grating coupler for a reference waveguide without QDPDs. The WG-QDPD showed a dark current of 106 nA at -3 V, as illustrated in Fig. 3a. Considering a WG-QDPD width of 30 µm and a length of 200 µm, the dark current density (1.8 $mA/cm^2$) is in line with the measurements obtained on the surface-illuminated QDPDs (2.8 $mA/cm^2$), indicating that the integration process does not introduce spurious leakage paths. The responsivity reaches 1.3 A/W (EQE of 74.8%) at an optical power of 116 nW, as shown in Fig. 3b. The significantly higher EQE compared to surface-illuminated QDPDs is attributed to increased absorption during propagation within the waveguides. For WG-QDPDs, the Si and QD stack constitute the hybrid waveguide designed for the propagation of the fundamental TE mode. Light is absorbed via the evanescent tail of the optical mode during its propagation, as depicted in **Appendix D, Figure 11**. In contrast to surface-illuminated QDPDs, which require a thick QD absorption layer for efficient light absorption, the extended absorption path of guided light in WG-QDPDs ensures efficient absorption even with a thin layer of QD. The combination of low dark current and high responsivity leads to a low estimated noise equivalent power (NEP=$\sqrt{2I_{dark}q}/R$, where $q$ is the elementary charge, and $R$ is the responsivity) of 0.15 pW/$\sqrt{Hz}$, indicating that WG-QDPDs are well-suited for weak signal detection. The photo-current exhibits nonlinear growth with optical power, a phenomenon attributed to QDPD saturation due to high series resistance, as observed in prior studies[34].

The speed of the fabricated WG-QDPD was measured at 1.55 μm, exhibiting a bandwidth of 1.1 MHz, as shown in Fig. 3c. Under large reverse bias of -3V, the thin QD absorption layer can be considered completely depleted[41]. Therefore, the response time ($\tau_{res}$) of the photodetector is determined by the drift progress ($\tau_{drift}$) and the resistance-capacitance constant ($\tau_{RC}$) of the electrical circuit $\tau_{res} = \sqrt{\tau_{drift}^2 + (2.2\tau_{RC})^2}$.[42] The drift time $\tau_{drift}$ is governed by the carrier mobility ($\mu$) and the voltage drop ($V$) across the depleted film thickness ($d$), with $\tau_{drift} = \frac{d^2}{\mu V}$. The other term, $\tau_{RC} = RC$, where $R$ represents the series resistance and $C$ denotes the capacitance. For a fully depleted QDPD, we estimate the series resistance $R$ using the access resistance (ITO sheet resistance and contact resistance between n-contact and ZnO), measured to be approximately 100 Ω for our WG-QDPDs. The capacitance $C$ is estimated to be around 4.1 pF ($C = \frac{\varepsilon_0 \varepsilon_r A}{d}$, where $\varepsilon_0$ represents the vacuum permittivity, $\varepsilon_r = 15.5$ denotes the dielectric constant of the PbS-QD film[41], $A = 6000$ μm$^2$ is the area of QDPD and $d = 195$ nm is the thickness of the PbS QD film). The estimated $\tau_{RC}$ is 0.41 ns, corresponding to a BW of 388 MHz using the relationship: BW = $0.35/\tau_{res}$[42]. Considering a measured BW of 1.1 MHz, the speed of our WG-QDPD is likely drift-limited. Furthermore, we measured the BW of WG-QDPDs with areas of 6000 μm$^2$ and 600 μm$^2$, both exhibiting similar bandwidths (see **APPENDIX E Figure 13**). This area-insensitive BW is contrary to the scenario where $\tau_{RC}$ dominates the speed of QDPDs[23,41], considering a consistent series resistance and area-related capacitance. This drift-limit speed can be further supported by examining $\tau_{drift}$. The carrier mobility of the 2.1 μm PbS-TBAI absorption layer, extracted via field effect transistor (FET) measurements[41], is $(0.94 \pm 0.6) \times 10^{-4}$ cm$^{-2}$V$^{-1}$s$^{-1}$ for electrons and $(1.41 \pm 0.35) \times 10^{-4}$ cm$^{-2}$V$^{-1}$s$^{-1}$ for holes. Considering a thickness of 75 nm for the absorption layer and a voltage drop of 3V, the estimated $\tau_{drift}$ is around 0.2 μs, corresponding to a bandwidth of 1.75 MHz. The carrier transition time in the 1.3 μm PbS-TBAI and 0.94 μm PbS-EDT layers is disregarded due to the significantly faster carrier mobilities exceeding $1 \times 10^{-3}$ cm$^{-2}$V$^{-1}$s$^{-1}$ in both layers[41]. Although the bandwidth of our demonstrated WG-QDPDs may not be as high as that of III-V photodiodes, it meets the requirements of many sensing applications. Further improvement on the speed of the WG-QDPD relies on improved ligand exchange strategies to boost the mobility of the 2.1 μm PbS QD layer. Reducing the thickness of the QD film can also decrease $\tau_{drift}$. This strategy was recently adopted to achieve QDPDs with a response time of 4 ns[41]. More details on the characterization of WG-QDPDs can be found in **APPENDIX E.**

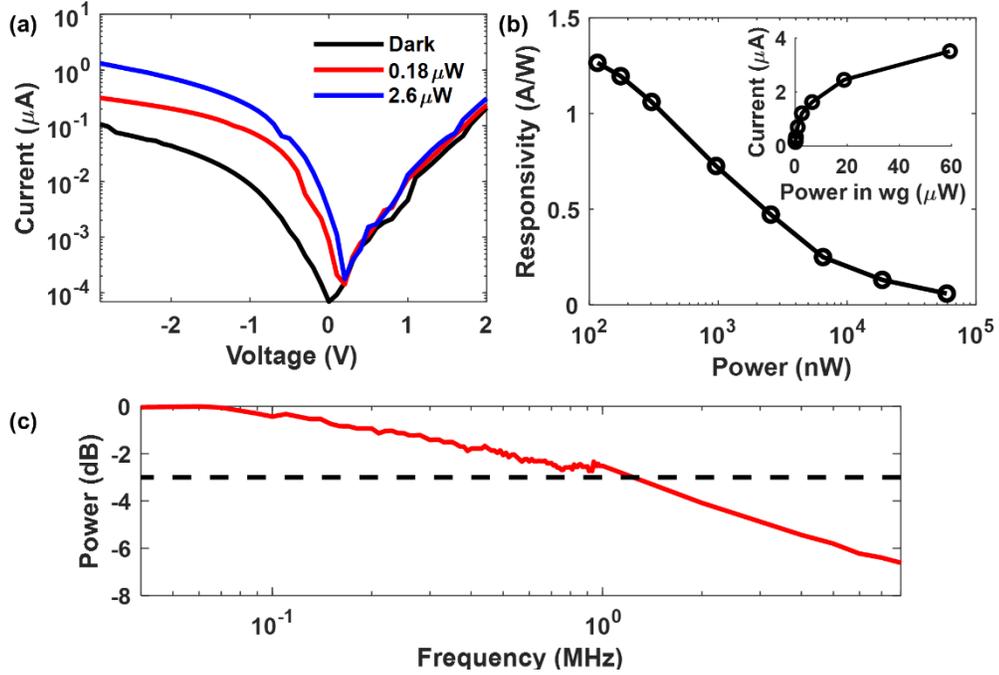

Fig. 3. Response of WG-QDPDs. (a) I-V curves of WG-QDPDs under dark condition and illumination at 2.1 μm. (b) Responsivity vs. optical power at -3 V bias voltage. Inset: photo-current vs. optical power. (c) Bandwidth of the integrated WG-QDPD.

On the same chip, we incorporated WG-QDPDs with varying lengths, as shown in Fig. 4a. We conducted transmission measurements on the waveguides after patterning each material. These measurements allowed us to ascertain the optical loss induced by each material, as shown in Fig. 4b. The ITO layer introduced an optical loss of around 0.02 dB/μm, resulting from strong free carrier absorption in the infrared[43]. The addition of the ZnO layer did not introduce measurable extra absorption. The entire QDPD stack raised the optical loss to 0.08 dB/μm, indicating the absorption is dominated by the QD film. This strong absorption competes with the absorption from the ITO bottom electrode, ensuring high responsivity of our WG-QDPDs. The dark current of these WG-QDPDs is proportional to the area, with a slope of 1.8 mA/cm$^2$, as shown in Fig. 4c. The responsivity of WG-QDPDs, measured at -3 V bias and 400 nW optical power, shows a saturation behavior with respect to PD length, as shown in Fig. 4d. The saturation behavior aligns with the prediction from optical loss measurements, as indicated by the blue dashed line in Fig. 4d.

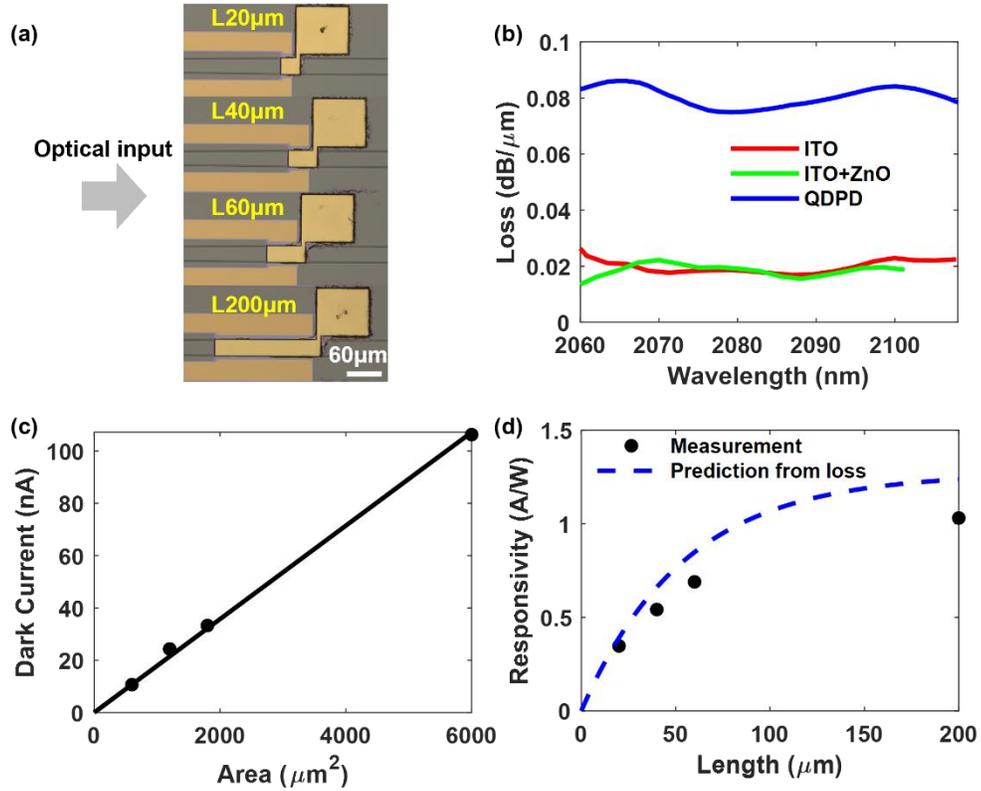

Fig. 4. Performance of WG-QDPDs with different lengths. (a) Top view of WG-QDPDs. (b) Optical absorption resulting from materials integrated onto waveguides. Measurements were obtained through optical transmission, with varying lengths for each material. (c) Darck current of devices with different area. (d) Responsivity of WG-QDPDs with different lengths. The predicted relationship between responsivity and length is presented by a dashed blue line, derived from optical loss measurements shown in Fig. 4b.

We also investigated WG-QDPDs with different shapes as shown in Fig. 5a. As we mentioned before, the waveguide width is extended to 30 µm to alleviate the optical power saturation, ensuring an efficient responsivity. However, this strategy also increases dark current due to the larger WG-QDPD area. Consequently, the NEP exhibits a trade-off between responsivity and dark current. Notably, heavy optical power saturation predominantly occurs in the initial section of the WG-QDPD. As light propagates within the waveguides, the optical power decreases exponentially. This insight allows us to narrow down the waveguide width at the far end without introducing significant power saturation. In our design, we implemented waveguide narrowing from 30 µm to 0.9 µm using both linear and exponential tapers, while keeping a consistent length of 200 µm. For the exponential taper, the width decreases at a rate of 0.023 µm$^{-1}$ (0.1 dB/µm), similar to the optical absorption measurement in Fig. 4b. The QDPD follows the same shape as these tapered waveguides. The linear taper reduces the WG-QDPD area to approximately 1/2, and the exponential taper further decreases the area to around 1/3. Consequently, the decrease in area reduces the dark current to 1/2 and 1/3, respectively, as shown in Fig. 5b. As expected, the responsivity of the tapered devices does not degrade significantly, as shown in Fig. 5c. This results in a champion NEP of 0.13 pW/√Hz for the exponentially tapered WG-QDPD, representing a 30% improvement compared to the NEP of the uniform WG-QDPD (0.18 pW/√Hz). Notably, there is an underestimation of the NEP here. The measurements were conducted at a relatively high optical power of 400 nW, which, combined with the nonlinear response to optical power, results in an underestimated responsivity.

We also fabricated a grating-assisted QDPD, where light injected from the left side was scattered vertically and absorbed by the QDPD, as shown in Fig. 5a. This configuration yielded the lowest dark current due to the smallest QDPD size of 30 µm by 30 µm. However, the responsivity was significantly lower compared to the evanescent coupling scheme. The degradation in responsivity is attributed to a combined effect of upwards coupling efficiency, incomplete optical absorption in the thin QDs layer, and possibly stronger power saturation.

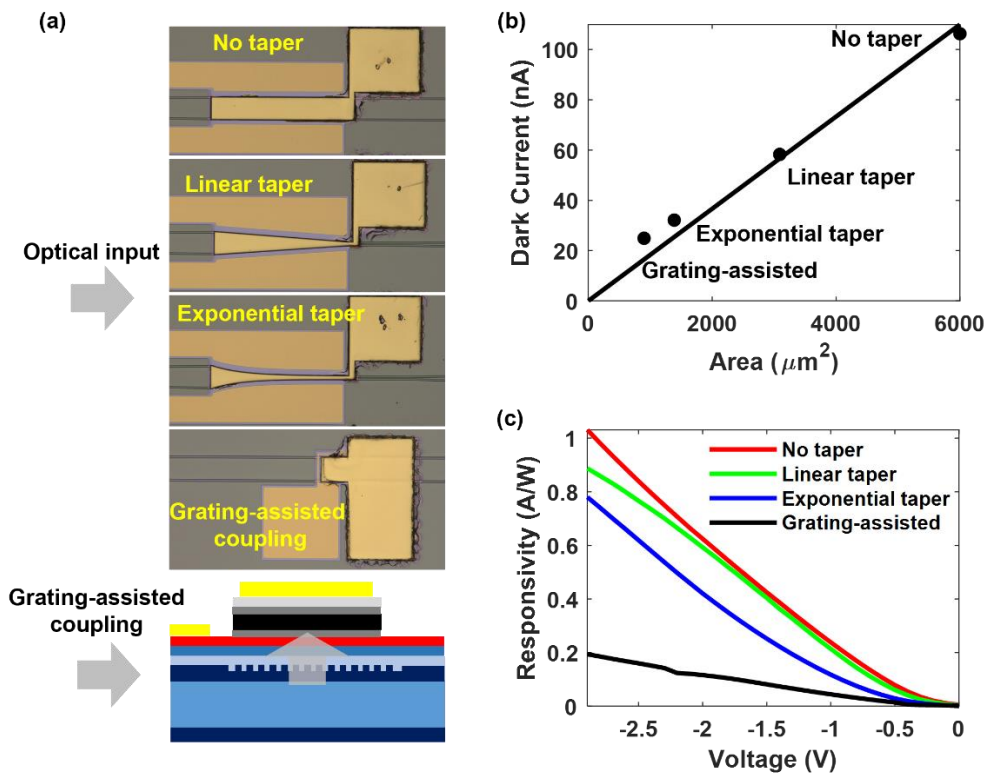

Fig. 5. Performance of WG-QDPDs with different shapes. (a) Top view of WG-QDPDs. Light is coupled into waveguides from the left side. (b) Dark current of different WG-QDPDs. (c) Responsivity of different WG-QDPDs.

*2.4 Demonstration of a 2.1 µm integrated spectrometer*

To demonstrate what complex tasks our proposed WG-QDPD integration approach can complete, we combined a WG-QDPD array with an eight-channel planar concave grating (PCG), forming a compact spectrometer, as shown in Fig. 6a. The PCG, designed with IPKISS[44], has a central wavelength of 2.1 µm and a wavelength range of around 70 nm. The fabrication was carried out on the same 220 nm SOI chip alongside the aforementioned components. To simplify the fabrication process, we used 70 nm shallow-etched distributed bragg reflectors (DBR) as retroreflectors[45]. FDTD simulations revealed that these DBR mirrors, with a period of 437 nm, exhibit reflection centered at 2.1 µm (>90%) and a 3 dB bandwidth of 200 nm, meeting the PCG requirements.

The transmission spectrum of the PCG was measured prior to QDPD integration, with a tunable laser around 2.1 µm (IPG Cr:ZnS/Se). The fabricated PCG covers a wavelength range from 2063 nm to 2135 nm, with a cross-talk better than -20 dB and insertion loss less than 3 dB, as shown in Fig. 6b, consistent with design metrics.

For the WG-QDPDs, we used a width of 30 µm and a length of 200 µm, to obtain an efficient photo response. After integrating the WG-QDPD array, the photo-response of each channel was measured at -1 V bias by sweeping the wavelength of the input laser. The response of the QDPD array aligns with the spectrum of the grating coupler (gray line in Fig. 6c), as measured from a reference WG-QDPD without the PCG. This consistent spectral shape indicates uniform responses of the WG-QDPDs, thereby suggesting its potential for scalable integration. The array of WG-QDPDs was measured directly without requiring further calibration. This type of spectrometer is interesting for sensing applications, such as glucose concentration monitoring.

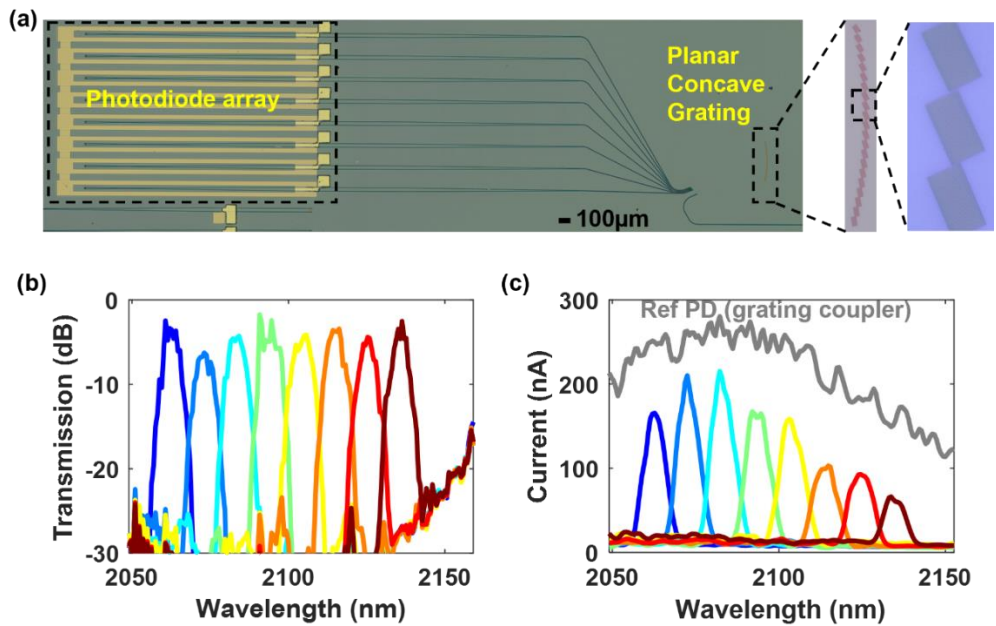

Fig. 6. Compact spectrometer based on a WG-QDPD array and a planar concave grating. (a) Top view of the spectrometer. Light is injected from the grating coupler on the right side. (b) Transmission spectrum of the planar concave grating. (c) Response of WG-QDPDs from eight channels and a reference waveguide (gray line).

*2.5 Discussion*

Compared with previously reported plasmonic HgTe QD photoconductors integrated on silicon waveguides[35], our integration approach achieves a similar level of dark current (106 nA), demonstrating more than one-order improvement in responsivity (1.3 A/W) and two-order improvement in bandwidth (1.1 MHz). When compared to alternative approaches for extending photodetection capabilities on the SOI platform, our proposed WG-QDPDs remain competitive in terms of dark current and responsivity. Notably, these results are on par with the best-performing bonded III-V photodetectors reported in the literature (responsivity 0.4-1.4 A/W around 2.3 µm, dark current 10 nA-2.5 µA) [6,8,9,46], outperforming monolithically grown GeSn (responsivity <0.52 A/W, dark current >1 mA)[13–16]. Another class of heterogeneous materials, 2D materials like graphene[47] or black phosphorus[48,49], also possess the potential to operate beyond the telecommunication wavelength. However, they face challenges in achieving high responsivity (0.07-0.3 A/W at wavelength of 2-3 µm)[11,12,50] and scalable integration. It is acknowledged that, despite advancements, the bandwidth of our WG-QDPD still lags behind compared to the GHz bandwidth demonstrated by other approaches[51–53]. Furthermore, additional efforts are necessary to mitigate the nonlinear response of our WG-QDPD.

The performance of the WG-QDPD could be further improved by optimizing the device structure. Previous reports on QDPDs at 2.1 µm have demonstrated an impressively low dark current density of $10^{-5}$ A/cm² and high responsivity[36]. Implementing such advancements could reduce the dark current of integrated photodetectors to sub-nanoampere levels, which is particularly advantageous for the detection of weak signals in sensing applications.

Also exploring alternative materials for long wavelength WG-QDPDs, such as HgTe[25,54,55] and $Ag_2Se$[26,27], is compelling to extend the photodetection capabilities within silicon photonics into the mid-infrared range, aligning with heightened interest in sensing applications. $Ag_2Se$ QDs, which rely on intra-band absorption, is thereby the more attractive material because of the greatly reduced toxicity. However, efforts are still needed to improve the responsivity and reduce the dark current for these materials.

**3. CONCLUSION**

In this work, we have demonstrated the integration of PbS QDPDs onto silicon waveguides, extending the photodetection capabilities beyond the traditional telecommunication range in silicon photonics. The achieved results at room temperature, including a low dark current of 106 nA, a high responsivity of 1.3 A/W at 2.1 µm, and a bandwidth of 1.1 MHz, highlight the effectiveness of our approach. The scalability of our integration method is exemplified through the presentation of an 8-channel compact spectrometer integrated with a WG-QDPD array. This integrated system offers a promising solution for on-chip spectroscopy around 2.1 µm. We believe that QDPD technology holds significant promise for cost-effective photodetection in the near and mid-infrared ranges in silicon photonics, especially for diverse sensing applications. Future work should focus on improving the QDPD characteristics, including detectivity, bandwidth and linearity, and extending sensing wavelength range.

**ACKNOWLEDGMENTS**

This work was supported by: European Research Council (ERC) under the innovation program grant agreement No. 884963 (ERC AdG NARIOS), FWO-Vlaanderen for research funding (FWO projects G0B2921N and G0C5723N), and China Scholarship Council (CSC Grant 201906120023).

**DATA AVAILABILITY**

Data available on request from the authors.

**APPENDIX**

**APPENDIX A: SYNTHESIS OF MATERIALS**

**Synthesis of 2.1 µm PbS QDs.** Lead oleate and N-n-hexyl-N'-dodecyl thiourea were prepared according to the procedure described by Hendricks et al.[37]. In a three-neck flask, 1.2 mmol lead oleate ( 0.9241 g) was dissolved in 20 mL n-dodecane at 150 °C under nitrogen atmosphere. Separately, N-n-hexyl-N'-dodecylthiourea (1 mmol, 0.3286 g) was mixed with 1 mL of diglyme (Diethylene Glycol Dimethyl Ether) in a vial and heated to 150 °C. The pre-heated thiourea solution was then injected swiftly into the lead oleate solution via a syringe and the reaction was allowed to run for 20 minutes at 150 °C. Subsequently, the flask was cooled down to ambient temperature by immersion into a water bath. The final dispersion underwent four purification cycles inside a nitrogen-filled glovebox, using a solvent mixture of 4:1 toluene to hexane, with methyl acetate as the non-solvent. The purified PbS QDs were dispersed in anhydrous n-octane to achieve the desired concentration.

**Synthesis of 0.94 μm PbS QDs.** Lead oleate and N-(3,5-bis(trifluoromethylphenyl))-N'-phenylthiourea were synthesized according to Hendricks et al.[37]. In a three-neck flask, 7.00 mmol lead oleate (5.38 g) was dissolved in 25 mL anhydrous n-octane at 90 °C under a nitrogen atmosphere. In a vial, 4.67 mmol of N-(3,5-bis(trifluoromethylphenyl))-N'-phenylthiourea (1,70 g, 1 eq.) and 2 mL of 1-methoxy-2-(2-methoxyethoxy)ethaan were mixed and heated to 90 °C as well. The thiourea solution was then quickly injected into the lead oleate solution via a syringe and the flask was cooled down to room temperature by immersion into a water batch after one minute. The resulting dispersion was purified four times by aid of n-octane and acetone and stored in anhydrous n-octane for further use.

**Synthesis of 1.3 μm PbS QDs.** In a three-neck flask, 2 mmol lead oleate (1.54 g) was dissolved in 20 mL n-dodecane at 120 °C under a nitrogen atmosphere, and flushed for 30 min. Next, 1.5 mmol of N-(p-(trifluoro-methyl)phenyl)-N'-dodecylthiourea (0.5828 g) was mixed with 1 mL of 1-methoxy-2-(2-methoxyethoxy)ethaan or diglyme under nitrogen atmosphere, preheated at 120 °C and quickly injected in the lead oleate solution at 120 °C. The reaction mixture was cooled down to room temperature by immersing the flask in a water bath after 80 sec. The resulting dispersion was then purified at least four times by aid of n-octane and acetone and stored in anhydrous n-octane for further use.

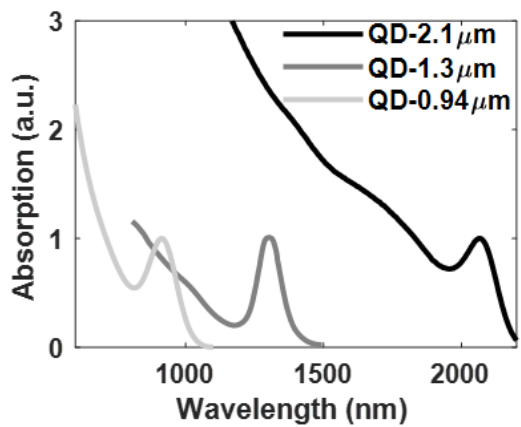

Fig. 7. Absorption spectra for three types of QD inks (dispersed in n-octane) used in QPDDs.

**APPENDIX B: DARK CURRENT SUPPRESSION**

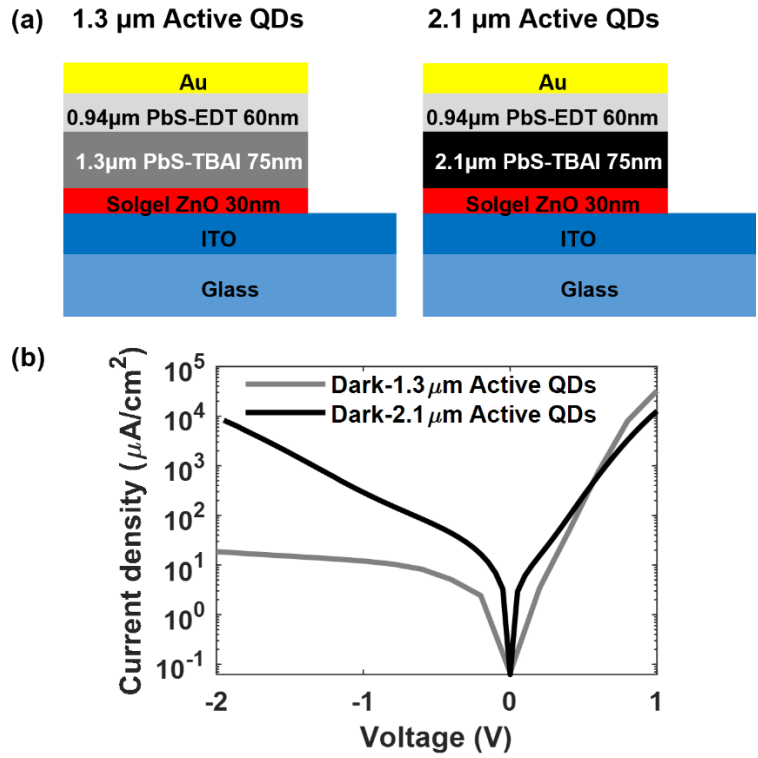

Fig. 8. Structure and characterization of QDPD with 1.3 µm and 2.1 µm PbS active layers. (a) QDPD structures. (b) Dark current density.

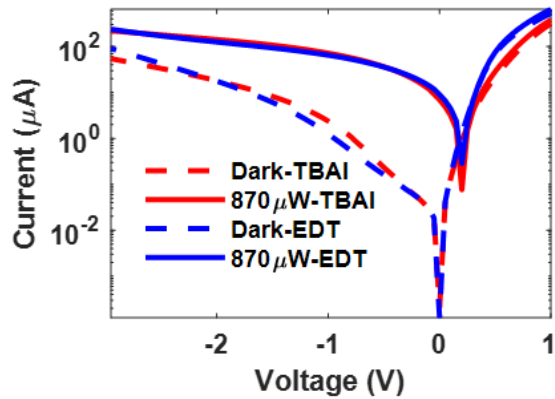

Fig. 9. Dark current and photocurrent of QDPD with an area of 1.77 mm². The photocurrent was measured with 2.1 µm laser with a gaussian beam illuminating the sample from the bottom glass side, with a peak power density of 220 mW/cm². Red: ZnO / 1.3 µm-PbS-

TBAI / 2.1 μm-PbS-TBAI / 1.3 μm-PbS-TBAI / 0.94 μm-PbS-EDT. Blue: ZnO / 1.3 μm-PbS-TBAI / 2.1 μm-PbS-TBAI / 1.3 μm-PbS-EDT / 0.94 μm-PbS-EDT.

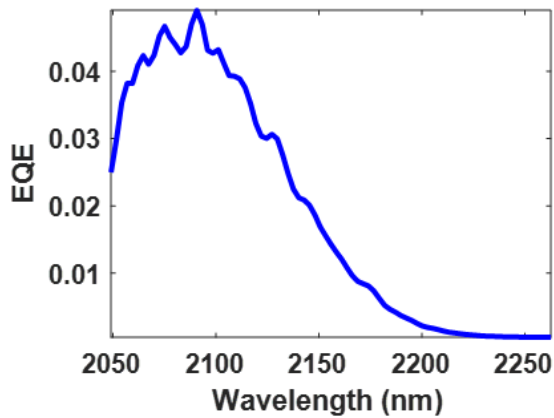

Fig. 10. Spectral response of the QDPD with structure B, measured at a bias of -1 V and a peak power density of 220 mW/cm$^2$.

**APPENDIX C: QDPD FABRICATION**

Substrates with 100 nm ITO on glass (2 cm × 1.5 cm, Ossila) were cleaned in acetone and isopropanol with sonication, followed by exposure to O$_2$ plasma (PVA TePla 600) for 10 min. A 30 nm layer of ZnO was then formed on top of the ITO layer via spin-coating a precursor solution and annealing at 325 °C for 30 minutes in ambient air. Following this, PbS QDs were spin-coated onto the ZnO layer, with subsequent treatment involving either 10 mg/mL tetra-n-butylammonium iodide (TBAI) in methanol or 0.01 vol% ethanedithiol (EDT) in methanol. The ligand exchange process for each cycle lasted 30 seconds, followed by two washes with methanol. Each cycle resulted in a film with a thickness of 25 nm for 2.1 μm PbS-TBAI, 30 nm for 1.3 μm PbS-TBAI, and 30 nm for 0.94 μm PbS-EDT. The QDs and ligand exchange steps were repeated multiple times to achieve the desired thickness. Finally, an 80 nm layer of gold was evaporated on top using a shadow mask, forming QDPD pixels with a diameter of 1.5 mm.

**APPENDIX D: WG-QDPD FABRICATION**

**Waveguides fabrication.** Waveguides were patterned with E-beam lithography (EBL) and dry etching. We used standard SOI substrates with 220 nm Si on top of 2 μm buried oxide in this work. 400 nm ARP6200.13 e-beam resist was used as a mask, exposed by EBL (Voyager Raith 50 kV) with a dose of 160 μC/cm$^2$ and developed in n-amyl acetate for 60 s. The resist pattern was transferred to Si by reactive ion etching (ICP-RIE, Oxford Instruments) using SF$_6$ and CHF$_3$ chemistry. A 70 nm shallow etch in Si was used to define the optical structures. We then spin-coated the flowable oxide hydrogen silsesquioxane (HSQ, Dow FOX-15) on top of the sample as a top cladding layer. Diluted HSQ (HSQ : MIBK = 1:4) was spun with a speed of 4000 rpm and an acceleration of 1000 rpm/s. The sample was then cured at 400 °C for 2 hours in a nitrogen atmosphere, achieving a cladding thickness of approximately 45 nm.

**ITO deposition and patterning.** 18 nm thick ITO was deposited on the sample by magnetron sputtering. Then the ITO film was patterned with photolithography (AZ5214 E) and wet etching (37% HCl:H2O=1:5, 15s).

**ZnO deposition and patterning.** 30 nm thick ZnO was deposited on the sample using the same solgel method outlined in APPENDIX C. Subsequently, the ZnO film underwent patterning through photolithography (AZ5214 E) followed by wet etching (37% HCl:H2O = 1:1000, 5s).

**N-contact deposition and patterning.** The N-contact was patterned using photolithography and the liftoff method. The desired pattern was generated using AZ5214 E photoresist, followed by metal evaporation with a composition of 20 nm Ti and 100 nm Au. The sacrificial photoresist and the metal layer on top were removed by immersing the sample in acetone for 1 hour.

**QDs deposition and patterning.** The QD film was patterned using EBL and the liftoff method. The desired pattern was generated using 400 nm thick ARP672.08 e-beam resist and EBL. Then QD stacks were spin-coated on top with the same procedure in APPENDIX C. The sacrificial e-beam resist and QDs on top were removed by immersing the sample in acetone for 10 min. PMMA has good chemical compatibility with the solvent involved (n-octane and methanol) in QDs deposition.

**P-contact deposition and patterning.** The P-contact was patterned using EBL and the liftoff method. The desired pattern was generated using 1 µm thick ARP672.08 e-beam resist and EBL. In this process, the e-beam resist baking was adjusted to 60 °C for 5 min on the hotplate, deviating from the standard 150 °C baking temperature. This modification aimed at protecting QDs and avoiding dark current degradation. A short sonication step was incorporated at the end of e-beam resist development to ensure a clean interface between QDs and the P-contact metal. Subsequently, an 80 nm thick layer of Au was evaporated as the P-contact metal. The sacrificial e-beam resist and the layer of QDs on top were removed by immersing the sample in acetone for 30 minutes.

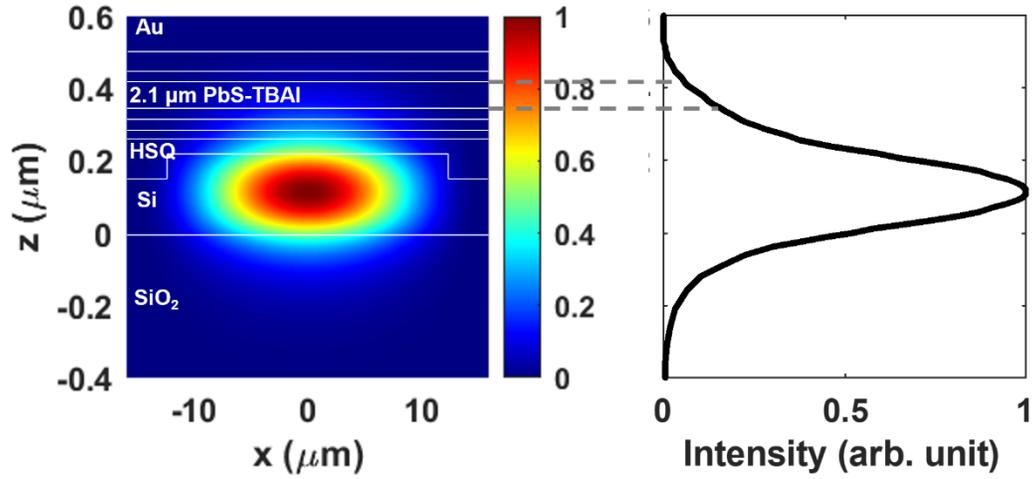

Fig. 11. Intensity profile of the fundamental TE mode in the WG-QDPD. The cross-section at position x=0 is shown in the right figure. Light is absorbed through the evanescent tail of the optical mode extended into the 2.1 µm QD film.

**APPENDIX E: WG-QDPD CHARACTERIZATION**

For DC measurements, a 2.1 µm laser ($Cr^{2+}$:ZnS laser, IPG Photonics) was coupled to a ZrF4 fiber and then further coupled to the WG-QDPD through a grating coupler. I-V curves were characterized with a source measure unit (Keithley 2400).

For AC measurements, the setup, as illustrated in Figure 12, involved a 1.55 µm laser (Santec 510) that was modulated by a Mach-Zehnder electro-optical modulator (EOM). The EOM was adjusted to its linear operation point, where applied radio frequency signals (SMT03, Rohde & Schwarz) were converted to optical intensity modulation. The modulated optical signals were amplified by Erbium-doped fiber amplifier (EDFA) and then coupled to the WG-QDPD through edge coupling. Polarization controllers (PC) were employed before the EOM and our chip to achieve the desired polarization. A DC bias was applied to the WG-QDPD using a source measure unit (Keithley 2400). The photodetector signal was collected by an electrical spectrum analyzer (ESA, FSP, Rohde & Schwarz). Both the DC bias and ESA were connected to the WG-QDPD with a bias tee.

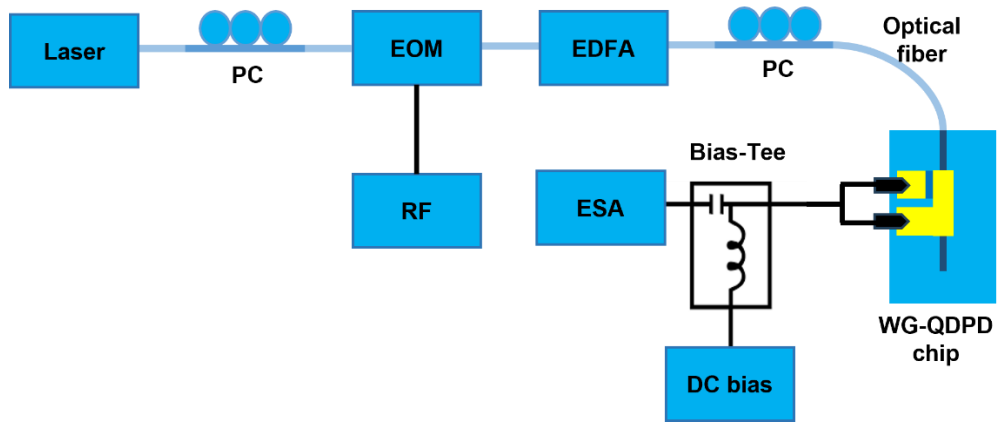

Fig. 12. Setup schematics used for bandwidth measurement. PC: polarization controller, EOM: electro-optical modulator, EDFA: Erbium-doped fiber amplifier, RF: radio-frequency source, ESA: electrical spectrum analyzer.

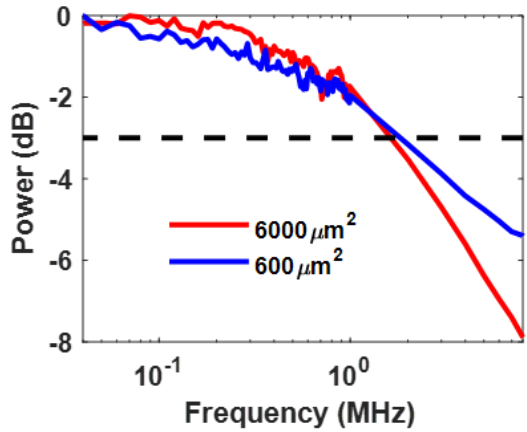

Fig. 13. Bandwidth of WG-QDPDs with different areas. WG-QDPDs, with a dimension of 30 µm × 200 µm and 3 µm × 200 µm, were measured under a reverse bias of -5 V.